\documentclass[preprint,prd,floatfix,preprintnumbers]{revtex4}
\usepackage{graphicx}

\begin{document}

\preprint{NT@UW-03-04, JLAB-THY-03-28, ADP-03-110/T548}

\title{Parton distribution functions in the pion from lattice QCD}

\author{W.~Detmold} 
\affiliation{Department of Physics, University of
  Washington, Box 351560, Seattle, WA 98195, U.S.A.}
\author{W.~Melnitchouk} 
\affiliation{Jefferson Lab, 12000 Jefferson
  Avenue, Newport News, VA 23606, U.S.A.}  
\author{A.~W.~Thomas}
\affiliation{Department of Physics and Mathematical Physics and
  \\ Special Research Centre for the Subatomic Structure of Matter,
  \\ University of Adelaide, Adelaide, SA 5005, Australia.}

\begin{abstract}
  We analyze the moments of parton distribution functions in the pion
  calculated in lattice QCD, paying particular attention to their
  chiral extrapolation.  Using the lowest three non-trivial moments
  calculated on the lattice, we assess the accuracy with which the $x$
  dependence of both the valence and sea quark distributions in the
  pion can be extracted. The resulting valence quark distributions at
  the physical pion mass are in fair agreement with existing Drell-Yan
  data, but the statistical errors are such that one cannot yet
  confirm (or rule out) the large-$x$ behavior expected from hadron
  helicity conservation in perturbative QCD. However, one can expect
  that the next generation of calculations in lattice QCD will allow
  one to extract parton distributions with a level of accuracy
  comparable with current experiments.
\end{abstract}  

\maketitle

\section{Introduction}
\label{sec:intro}

It is widely appreciated that the pion plays a very fundamental role
in QCD. Given that chiral symmetry is such a good symmetry of nature,
because of the extremely low masses of the $u$ and $d$ quarks, the
pseudo-Goldstone character of the pion is ubiquitous in hadron
physics.  As a result, the determination of its structure, both from
experiment and non-perturbative studies of QCD, is of great
importance.
The parton distribution
functions (PDFs) of the pion have been measured in a number of
experiments, using the Drell-Yan reaction
\cite{Badier:1983mj,Betev:1985pg,Aurenche:1989sx,Bonesini:1987mq,Conway:fs}.
Such experiments tend to focus on the region of Bjorken-$x$ above
$\approx 0.2$ and hence are most sensitive to the valence
distribution. Until recently there was little constraint on the size
or form of the sea quark distributions, but measurements of
charge-exchange in semi-inclusive 
deep inelastic scattering (DIS) 
at HERA have yielded some
information at very low $x$
\cite{Klasen:2001dd,Holtmann:1994rs,Levman2001}, and one can also
expect new, high precision data from semi-inclusive DIS after the
upgrade at Jefferson Lab \cite{JLAB12}. This observation will be
important for our analysis because the current errors for the sea
quark distributions are considerably larger than the statistical
errors in the first moment of the lattice data.
  
The existing data have been used to constrain various phenomenological
parameterizations of the pion PDFs
\cite{Owens:1984zj,Sutton:1991ay,Gluck:1999xe,Gluck:1991ey,Gluck:1997ww}.
At the same time they are used to guide and test non-perturbative
models of the internal structure of the pion, from the constituent
quark model \cite{Altarelli:1995mu} to the NJL model
\cite{Bentz:1999gx,Shigetani:dx,Davidson:2001cc,RuizArriola:2002bp}
and others
\cite{Dorokhov:gu,Bissey:2002yr,Frederico:dx,Melnitchouk:2002gh}.  In
addition, there has recently been a calculation within a covariant
model, based on a truncation of the Dyson-Schwinger equations
\cite{Hecht:2000xa}.

One of the clearest predictions for the $x$ dependence of the pion
structure function comes from considerations of hadron helicity
conservation within perturbative QCD
\cite{Farrar:yb,Farrar:1979aw,Lepage:1980fj,GUNION}.  It is a firm
expectation within this framework that the valence quark distribution
should behave like $(1-x)^2$ as $x \rightarrow 1$.  On the other hand,
the experimental data seem to be more consistent with a form linear in
$(1-x)$.  One suggestion is that the experimental data may have a
substantial higher-twist component \cite{BB}.  We shall see that the
analysis of data from lattice QCD offers a significant possibility of
resolving the issue in the near future.

In Section~\ref{sec:lattice} we review the lattice simulations of the
moments of the pion structure function, while the chiral extrapolation
of these moments is described in Section~\ref{sec:xtrap}.  The
reconstruction of the $x$ dependence of the valence and sea quark
distribution functions in the pion is presented in
Section~\ref{sec:reconstruct}. In order to make quite clear what can
be learned from existing lattice data, and what might become possible
in the near future, we present several alternative methods for
performing the extraction. In Section~IV, we also investigate the
pion mass dependence of the reconstructed distribution. Finally, in
Section~\ref{sec:summary} we summarize our results, and outline future
applications of the methodology presented here.

\section{Lattice Results}
\label{sec:lattice}

By discretizing space-time as a four-dimensional hyper-cubic lattice,
the field equations of QCD can be solved numerically in the
non-perturbative region.  The potential of lattice QCD is that it
allows a first principles investigation of hadron properties and
structure.  The main weakness of these numerical calculations is the
vast computational resources that they require.  Indeed, it is not yet
computationally feasible to perform lattice calculations that
correspond to the parameters of the real world.  Current simulations
are run at quark masses 3--10 times too large, on lattice volumes that
are likely too small, and often use the quenched approximation (in
which sea quark loops are neglected).  The result of these
restrictions is that various extrapolations are necessary to reach the
physical regime.

The pioneering lattice calculations of hadron structure functions were
made by Martinelli and Sachrajda in the late 1980s
\cite{Martinelli:1987bh,Martinelli:1987zd}. Even though the available
computational resources restricted the statistical accuracy of their
studies and confined them to small lattices, their results are still
consistent with the more advanced calculations of the QCDSF
collaboration which we discuss below.  First, however, we briefly
consider the formalism needed to connect the lattice and continuum
theories.

While the $x$ dependence of the parton distribution functions cannot
be computed directly on the lattice, one can compute the moments,
$\langle x^n \rangle$, of the distributions.
Using the operator product expansion, these moments can be related to
matrix elements of operators of a given twist.  The leading twist
(twist-2) operators are given by
\begin{eqnarray}
{\cal O}^{\mu_1 \ldots \mu_n}_q &=&
  i^{n-1}\ \overline{\psi}_q\
  \gamma^{\{\mu_1}\, D^{\mu_2} \cdots \,D^{\mu_n\}} \psi_q\ ,
\label{eq:operators}
\end{eqnarray}
where $\psi_q$ are quark fields, $D^{\mu}$ is the covariant
derivative, and the braces $\{ \cdots \}$ denote symmetrization of
indices.
For reference, we shall work with the $u$ quark distribution in the
$\pi^+$ meson, $u_{\pi^+}(x)$, which can be related to distributions
in the $\pi^-$ and $\pi^0$ by charge symmetry (c.f.
Refs.~\cite{Londergan:1998ai,Londergan:wp}),
\begin{eqnarray}
u_{\pi^+}(x)
&=& \bar d_{\pi^+}(x)\ =\ d_{\pi^-}(x)\ =\ \cdots       \nonumber\\
&\equiv& q_\pi(x)\ ,
\end{eqnarray}
where we have suppressed the dependence on the scale $Q^2$.  The
moments of $q_\pi(x)$ are defined as
\begin{eqnarray}
\label{eq:moments}
\langle x^n \rangle_q
&=& \int_0^1 dx\ x^n\ \left( q_\pi(x) - (-1)^n \bar q_\pi(x) \right)\ ,
\end{eqnarray}
where, for example, the $n=0$ moment corresponds to the number sum
rule, $\langle x^0 \rangle_q = 1$.  Operationally, these moments can
be extracted from the forward matrix elements of the operators
(\ref{eq:operators}) as
\begin{eqnarray}
\langle \pi(\vec p) |
        {\cal O}^{\mu_1 \ldots \mu_{n+1}}_q - {\rm traces}
| \pi(\vec p) \rangle
&=& \langle x^n \rangle_q p^{\mu_1} \cdots p^{\mu_{n+1}} \ ,
\end{eqnarray}
where ``traces'' are subtracted to give matrix elements that transform
irreducibly.

In the lattice formulation, discretized versions of the operators
(\ref{eq:operators}) must be defined that have the correct continuum
limits. A number of technical considerations arise in this procedure.
For the $n=1$ moment, there are two possible lattice discretizations
of the corresponding continuum operator.  One of these can be
evaluated with both pion states having zero momentum, which results in
greater statistical precision.  We only include the results for this
operator in our analysis.  The lattice data for the less well
determined operator are consistent with this, however, and their
inclusion would not modify our conclusions. For $n=2$ and $3$, non-zero
momentum is unavoidable and the data are correspondingly less precise.
Also, the reduced symmetry of the lattice means that it becomes
impossible to define operators that transform irreducibly for $n\ge4$.
Calculation of the corresponding moments is more difficult as it
necessarily involves the evaluation of coefficients which describe the
mixing with lower dimensional operators.  Consequently there are only
data for $n = 1, 2$ and 3 at the present time.

Although somewhat easier to calculate than for the case of the
nucleon, the moments of the pion distribution functions have received
less attention in the literature.  The QCDSF collaboration has
performed the only detailed study \cite{Best:1997qp} of the moments of
the pion parton distributions.  The analysis was based on a sample of
500 configurations, with the simulations performed in the quenched
approximation using a Wilson quark action at three different quark
masses, $m_q \simeq 70$, 130 and 190~MeV, on a $16^3\times 32$
lattice at $\beta=6.0$.  QCDSF set the scale by linearly extrapolating
the $\rho$ meson mass to the chiral limit.  Although there are
considerable uncertainties associated with such an extrapolation,
given the potential non-linearities associated with chiral
non-analytic behavior, the study by Leinweber {\it et al.}
\cite{Leinweber:2001ac} suggests that a linear approximation may not
be so inaccurate in this particular observable. In any case, with the
physical scale set in this way the lattice moments correspond roughly
to a scale $Q^2 \approx (2.4\ {\rm GeV})^2 \sim 5$--6~GeV$^2$
\cite{Best:1997qp}.  The QCDSF collaboration have also analyzed some
higher twist contributions to the pion structure
function \cite{Capitani:1999rv}, finding that they are rather small
(at least at the large quark masses considered).

While the QCDSF investigations used quenched field configurations, one
would expect that the effects of that approximation should be
relatively small at the large quark masses for which data are
available.  Indeed, previous comparisons of quenched and unquenched
data for nucleon structure calculations
\cite{Detmold:2001jb,Dolgov:2002zm} showed no statistically
significant difference in this region.  The QCDSF lattice study of
hadron structure is ongoing and we look forward to unquenched results
in the near future.  When lattice calculations are able to be
performed at significantly lighter masses, the effects of quenching
will become apparent.  {}Finally, we note that the lattice results
cannot be regarded as definitive, even at the masses used, until a
thorough investigation of the effects of the finite lattice spacing
and finite lattice volume has been undertaken.  {}For example, Jansen
\cite{Jansen:2000xm} suggests that the ${\cal O}(a)$ errors could be
significant in calculations of $\langle x\rangle_q$ with Wilson
fermions.  Bearing these caveats in mind, we take the lattice results
at face value in the current study, with the understanding that our
analysis can easily be updated to reflect improvements in the lattice
data as they occur.

We stress that, even though lattice QCD calculations in the next few
years will be extended to smaller quark masses and larger lattices in
the quenched and unquenched (or at least partially quenched
\cite{Sharpe:1997by}) versions of QCD, the numerical challenge of
light quark masses is such that extrapolation over a fairly large
range of quark mass will be needed for many years.

\newpage

\section{Chiral extrapolation}
\label{sec:xtrap}

The approximate chiral symmetry of QCD leads to the appearance of
pseudoscalar (Goldstone) bosons.  In the case of chiral
SU(2)$_L\times$SU(2)$_R$, these are identified with the pions,
$\pi^{\pm,0}$.  Because the pion mass vanishes with the square root of
the current quark mass, $m_\pi \sim \sqrt{m_q}$, the pion takes on an
increasingly important role in QCD as $m_q \to 0$.  Its effect on
hadron structure can be quantified using systematic expansions of
observables in powers (and logarithms) of $m_\pi$ \cite{ChiPT}.  In
particular, because of the structure of the Goldstone boson loop
corrections to hadronic properties, coefficients of terms in the
expansions which are non-analytic in the quark mass can be calculated
in terms of physical parameters, and hence are model independent.  For
the case of the nucleon, the leading non-analytic behavior of the
moments of parton distribution functions arising from such loops was
found to be crucial in understanding the relation between the lattice
results and the physical values of the
moments~\cite{Detmold:2001jb,Thomas:2000ny,Detmold:2001dv,Detmold:rw,Detmold:2002ac,Detmold:2002nf,Arndt:2001ye,Chen:2001eg}.
Any serious extrapolation of lattice calculations from the
unphysically large quark masses at which they are currently performed
to the physical quark masses must incorporate the effects of the pion
cloud \cite{Thomas:2002sj}.

Arndt and Savage \cite{Arndt:2001ye} have calculated the leading
chiral corrections to the moments of the pion's quark distributions,
finding that the pion cloud contributions to the $C$-odd ($n$-even)
flavor non-singlet (NS) moments receive corrections:
\begin{equation}
\langle x^n\rangle_q^{\rm NS}
= a_n
  \left[ 1-\frac{1-\delta^{n0}}{(4\pi f_\pi)^2}
         m_\pi^2 \log\frac{m_\pi^2}{\Lambda_\chi^2}
  \right]\ ,
\end{equation}
where $f_\pi=93$~MeV is the pion decay constant, $a_n$ is the value of
the moment in the chiral limit and $\Lambda_\chi \sim 4 \pi f_\pi
\approx 1$~GeV is the chiral scale.  The $n=0$ moment is not
renormalized by pion loops because of charge conservation.  In the
singlet sector, for the $C$-even ($n$-odd) moments, pion loops do not
introduce any non-analytic structure. Physically, this is because any
momentum lost by valence quarks through pion emission is recovered
through the additional sea quarks generated.  Of course, the $C$-even
non-singlet and $C$-odd singlet moments must vanish identically
because of the crossing symmetry properties of the distributions.

Since the lattice data for the moments of the pion PDF are well fit by
a linear function of $m_\pi^2$, over the region where they have been
calculated, it is natural to apply a functional form similar to that
used to extrapolate the moments of the NS PDF of the nucleon
\cite{Detmold:2001jb}.  We modify the linear term only minimally,
replacing $m_\pi^2$ by $m_\pi^2/(m_\pi^2 + M^2)$ so that this term
goes to a constant, rather than diverging, as $m_\pi \rightarrow
\infty$,
\begin{eqnarray}
\label{eq:xtrap_NS}
\langle x^n \rangle_q^{\rm NS}
&=&  a_n
\left[ 1 - c_{\rm LNA}
  m_\pi^2\log\left( \frac{m_\pi^2}{m_\pi^2+\mu^2} \right)
\right]
+ \frac{b_n m_\pi^2}{m_\pi^2 + M^2}\ ,\ \ \ \ n > 0\ ,		\\
\label{eq:xtrap_S}
\langle x^n \rangle_q^{\rm S}
&=& \bar{a}_n + \bar{b}_n { m_\pi^2 \over m_\pi^2 + M^2 }\ ,
\end{eqnarray}
where $a_n, b_n, \bar{a}_n$ and $\bar{b}_n$ are fit parameters, and
$c_{\rm LNA} = 1/(4\pi f_\pi)^2$ is the model independent coefficient
of the leading non-analytic (LNA) term in the non-singlet expansion.
The fits are insensitive to the parameter $M$ as long as it is large,
and in this analysis it is fixed at $M=5$~GeV.

The behavior of the moments in the limit $m_q \to \infty$ can be
determined model independently from heavy quark effective theory, so a
more ambitious scheme would be to build this behavior into the fitting
function as well.  In the heavy quark regime, contributions from the
quark-antiquark sea are suppressed as $1/m_q^2$ and the two valence
quarks in the pion each carry half of the momentum of the pion.  The
corresponding valence distribution is therefore a $\delta$-function
located at $x=1/2$, so that the moments behave as
\begin{equation}
  \label{eq:HQL}
  \langle x^n \rangle_q \longrightarrow { 1 \over 2^n } \, ,
  \ \ \ \ m_q\to\infty\ .
\end{equation}
This limit is easily built into the (non-linear) extrapolations, along
with the chiral non-analytic behavior (see Ref.~\cite{Detmold:2001dv}
for the analogous case of the nucleon).
However, given the present accuracy of the lattice data it is
sufficient to use the simpler extrapolation functions, given in
Eqs.~(\ref{eq:xtrap_S}) and (\ref{eq:xtrap_NS}), which are not
constrained by the heavy quark limit.

The parameter $\mu$ in the argument of the chiral logarithm in
Eq.~(\ref{eq:xtrap_NS}) is physically related to the size of the source
of the pion cloud and controls the onset of the chiral behavior in the
NS moments as $m_\pi \to 0$. Ideally its value will be determined from
fits to unquenched lattice data, however, present data are not yet at
sufficiently low masses. Instead we take the value $\mu=0.7$~GeV,
which is somewhat larger than that used in the nucleon analysis
because of the smaller size of the pion, and test the sensitivity to
$\mu$ by varying it over the range $(0.4,1.0)$~GeV.

\begin{figure}[!t]
  \includegraphics[width=0.9\columnwidth]{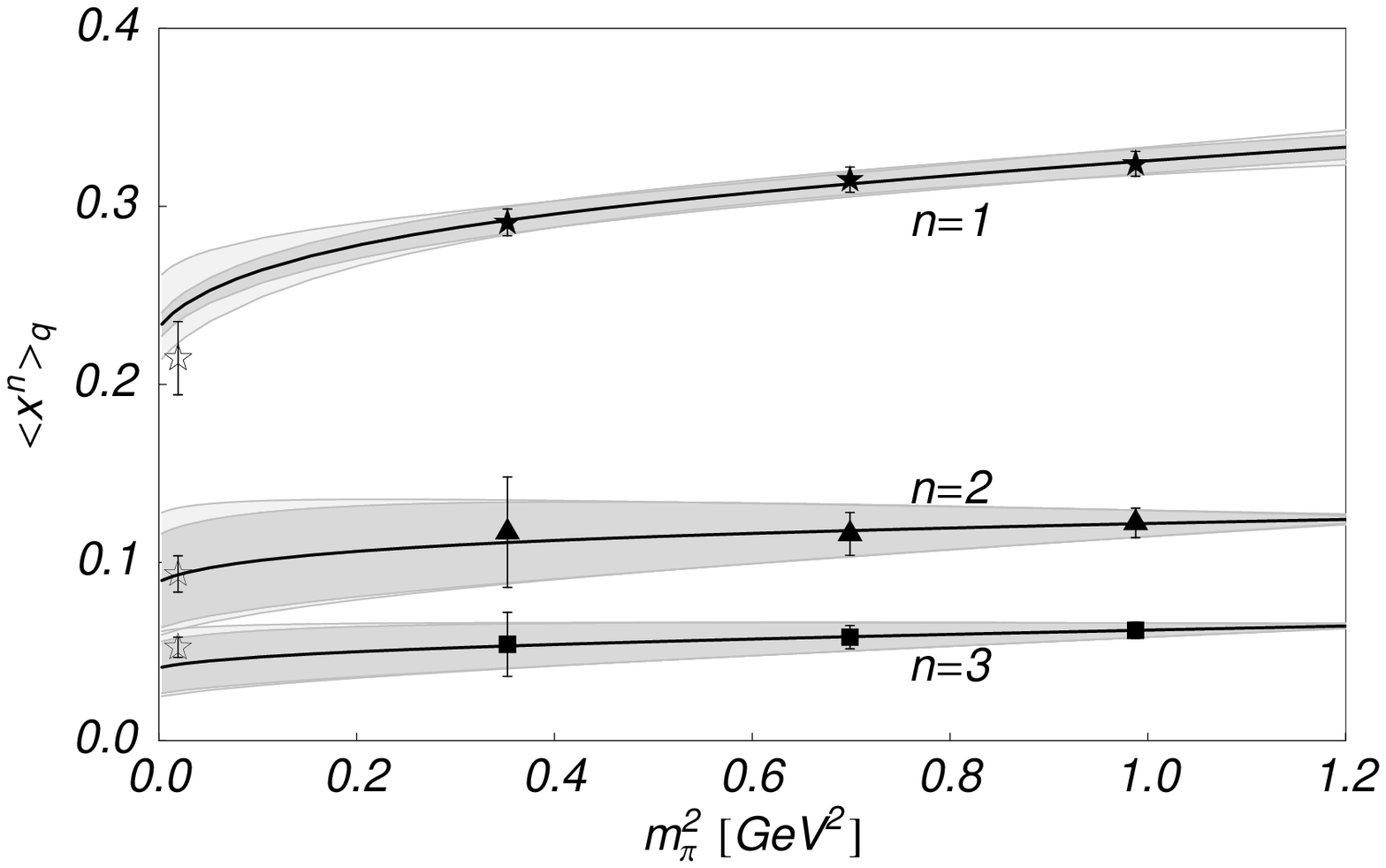}
  \includegraphics[width=0.9\columnwidth]{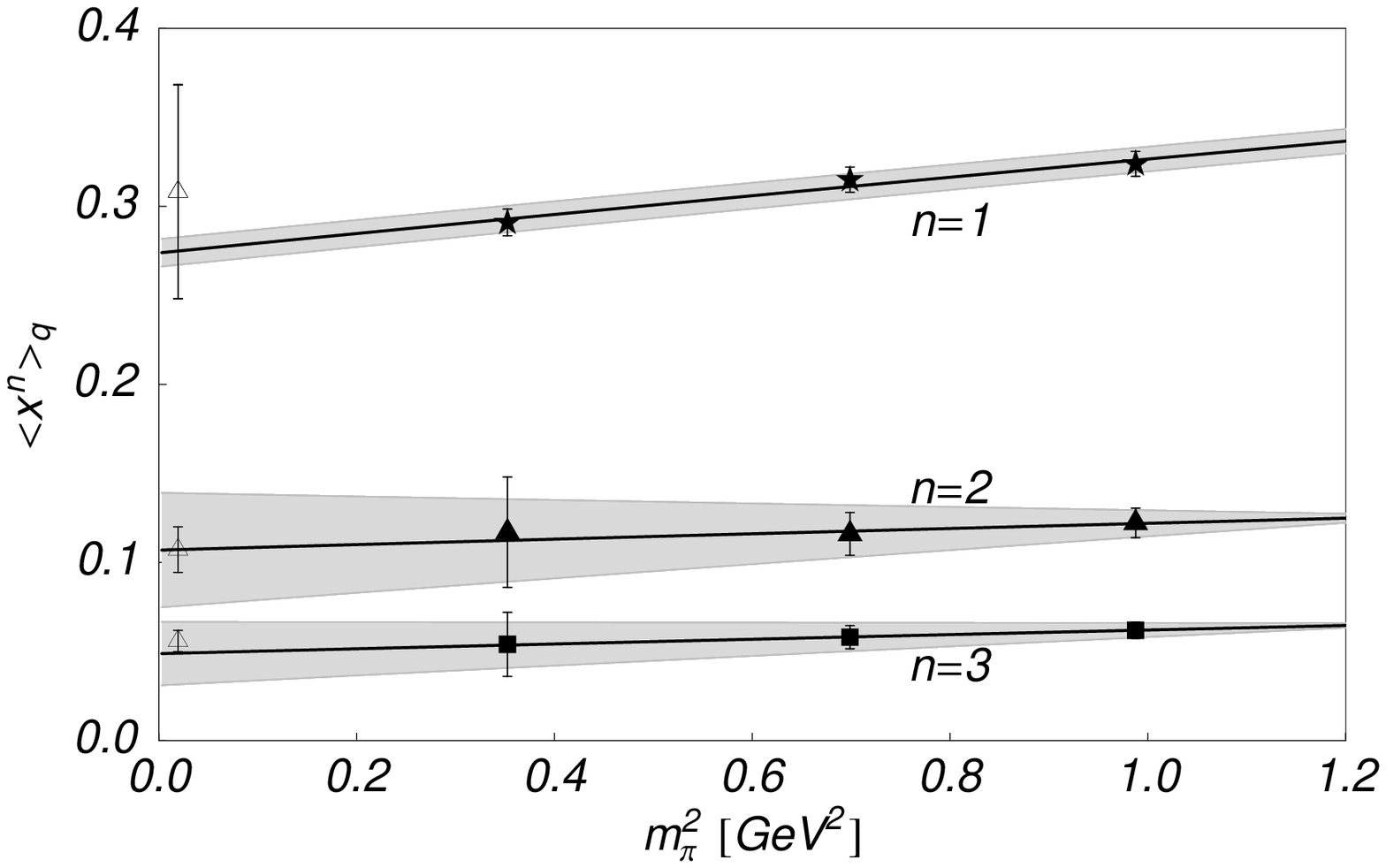}
\caption{\label{fig:xtrap}
  Chiral extrapolation of the lowest three lattice moments
  \protect\cite{Best:1997qp} of the pion distributions.  The upper
  plot shows the extrapolation of the {\em valence} moments and lower plot
  shows that of the {\em total} moments.  The solid curves correspond to
  fits using Eq.~(\ref{eq:xtrap_NS}), with $\mu=0.7$~GeV for the valence
  moments, and Eq.~(\ref{eq:xtrap_S}) for the singlet moments. The dark
  shaded region in both plots corresponds to fits to the data plus or
  minus their error bars, while the lighter shaded regions in the valence
  plot show the additional effect of varying $\mu$ between $0.4$ and
  $1.0$~GeV on top of the statistical variation.  The phenomenological
  valence (open stars) and total (open triangles) moments are shown at
  $m_\pi^{\rm phys}$ (see Section~\ref{sec:reconstruct}).}
\end{figure}

The above results for the chiral extrapolation are valid in full QCD,
whereas the existing lattice data have been generated within the
quenched approximation (in which the effects of background quark loops
are neglected).  Because quark loop effects are proportional to
(powers of) $1/m_q$, one expects loops to play a relatively minor role
at large quark mass.  Indeed, for moments of the nucleon parton
distributions the quenched and full QCD simulations were found
\cite{Dolgov:2002zm} to be equivalent within statistical errors for
$m_\pi \agt 0.5$--0.6~GeV.  Therefore, in the present analysis we
assume that the available (quenched) data at large $m_\pi$ provide a
reliable estimate of the unquenched moments at $m_\pi \agt
0.5$--0.6~GeV. Future simulations will allow quantitative tests of
this assumption, and when they can be performed at quark masses light
enough for the difference to become apparent our analysis will need to
be repeated using quenched \cite{Chen:2001gr} (or partially quenched
\cite{Chen:2001yi}) chiral perturbation theory.

In general the matrix elements receive contributions from diagrams in
which the operator insertions are either on quark lines which are
connected to the pion source (CI), or on quark lines which are
disconnected (DI) ({\em i.e.} connected only through gluon lines to
the pion source):
\begin{eqnarray}
\langle x^n \rangle_q
&=& \langle x^n \rangle_q^{\rm CI} + \langle x^n \rangle_q^{\rm DI}\ .
\end{eqnarray}
The disconnected insertions contribute only to the singlet operators, 
while the connected diagrams contribute
in both the singlet and non-singlet cases. The evaluation of
disconnected diagrams is considerably more difficult numerically, and
thus far only the connected pieces, $\langle x^n \rangle_q^{\rm CI}$,
have been computed \cite{Best:1997qp}.
Once again we can make use of the large quark masses at which the
lattice moments have been simulated by noting that the disconnected
insertions should also be suppressed for $m_\pi \agt 0.5$~GeV, so
that
\begin{eqnarray}
\langle x^n \rangle_q
&\approx& \langle x^n \rangle_q^{\rm CI}\ ,\ \ \ \
 m_\pi \agt 0.5~{\rm GeV}\ .
\end{eqnarray}
This same argument also suggests that, {\it at these values of $m_\pi$},
the pion PDFs should satisfy
\begin{eqnarray}
q_{\rm total}(x) \equiv q_\pi(x) + \bar q_\pi(x)
&\approx& q_\pi(x) - \bar q_\pi(x)\ \equiv\ v_\pi(x)\ , 
\end{eqnarray}
where $v_\pi(x)$ is the valence quark
distribution in the pion.
This observation allows one to approximate the $C$-odd (valence)
moments by the $C$-even moments at large $m_\pi$, and extrapolate them
according to Eq.~(\ref{eq:xtrap_NS}) to compare with the
phenomenological valence (non-singlet) moments.  However, we stress
that future lattice data for the $n$-odd moments should vary smoothly
as $m_q$ decreases -- {\em i.e.} they should show no rapid,
non-analytic behavior as the chiral limit is approached.

In Fig.~\ref{fig:xtrap} we show the lattice data from the QCDSF
collaboration \cite{Best:1997qp} for the $n=1$, 2 and 3 moments (of
the $u$ quark distribution in the $\pi^+$) as a function of $m_\pi^2$.
The fits to the $n=2$ data using Eq.~(\ref{eq:xtrap_NS}) and those to
the $n=1$ and 3 moments using Eq.~(\ref{eq:xtrap_S}) are indicated by
the curves in the upper and lower plots, respectively. For each fit
the dark shaded error bands correspond to fits to the data $\pm$
errors.  The phenomenological values of the moments, indicated by open
stars (valence distribution) and open triangles (total distribution)
at $m_\pi^{\rm phys}$, are taken from an average of global fits
\cite{Sutton:1991ay,Gluck:1999xe} to the pion structure function data
(see Section~\ref{sec:phenom} below).  Assuming valence quark
dominance of the moments at $m_\pi \agt 0.5$~GeV, we also show in the
upper plot the $n=1$ and 3 moments extrapolated as if they were
non-singlet, using Eq.~(\ref{eq:xtrap_NS}).  Under the same assumption
we extrapolate the $n=2$ moment as if it were a singlet in the lower
plot.  In the central curves of the NS fits, we choose $\mu=0.7$~GeV.
The outer, lightly-shaded envelopes show a conservative variation of
this parameter between 0.4 and 1.0 GeV in addition to the statistical
variation (dark shaded region).  In all cases the extrapolated
moments, both singlet and non-singlet, agree with the phenomenological
values within errors, as shown in Table~\ref{tab:moment}.  This
provides {\em a posteriori} evidence for the valence dominance of the
moments (suppression of quark loops) at large $m_\pi$. 
\begin{table}
\begin{ruledtabular}\begin{tabular}{c|cccc}
$\langle x^n \rangle_q$ & $n=0$ & $n=1$ & $n=2$ & $n=3$ \\ \hline
\multicolumn{5}{c}{Moments of Phenomenological PDFs}    \\ \hline
valence & 1  & 0.21(2) & 0.09(1)  & 0.052(5)            \\
sea [Eq.~(\ref{eq:seadef})]   & -- & 0.05(3) & 0.007(4) & 0.002(1)       \\
total   & -- & 0.31(6) & 0.11(1)  & 0.056(6)            \\ \hline
\multicolumn{5}{c}{Extrapolated Lattice Moments}        \\ \hline
valence [method (ii)] & 1  & 0.24(1)(2) & 0.09(3)(1) & 0.043(15)(3)       \\
valence [method (iii)] & 1  & 0.18(6) & 0.10(3)(1) & 0.05(2)    \\
sea    & -- & 0.03(1) &  -- & 0.001(9)\\ 
total   & -- & 0.275(8) & 0.11(3) & 0.05(2)             \\
\end{tabular}\end{ruledtabular}
\vspace*{0.5cm}
\caption{\label{tab:moment}
  Moments of PDFs of the pion, obtained from phenomenological PDFs
  (see Section~\ref{sec:reconstruct}) and extrapolated from the lattice
  (as discussed in the text).
  The $n=2$ lattice total moment is obtained from the lattice
  valence moment by adding twice the phenomenological sea. The lattice sea is
  determined by subtracting the phenomenological valence
  moments from the lattice total moments (for $n$-odd).  
  Errors on the extrapolated lattice 
  moments are calculated from fits to the data $\pm$ their errors
  (first parentheses) and from varying the
  parameter $\mu$ between 0.4 and 1.0 GeV (second parentheses, 
  where applicable). } 
\end{table}
%

\section{Reconstruction of the quark distribution}
\label{sec:reconstruct}

In this section we use the available lattice moment data to constrain
the Bjorken-$x$ dependence of the underlying PDFs.  The approach
adopted here is similar to that in the earlier analysis of the PDFs in
the nucleon \cite{Detmold:2001dv,Detmold:rw,polrecon}.  The general
procedure is to choose a particular parameterization for the $x$
dependence of the distribution, and perform a Mellin transformation to
give the parametric dependence of its moments.  Values for the
moments, extrapolated from the lattice data, can then used to fit the
various parameters and reconstruct the physical distribution.

\subsection{Phenomenological distributions}
\label{sec:phenom}

Before using a specific parameterization to analyze the lattice data,
we first test the robustness of the procedure by examining the extent
to which the parameters of the phenomenological valence distributions
can be reconstructed from their lowest moments.  This will provide an
estimate of the systematic error in the choice of parameterization and
the reconstruction procedure.

Several groups have performed global analyses of pion structure
function data and constructed parameterizations of the parton
distributions.  The valence quark distribution in the SMRS
parameterization \cite{Sutton:1991ay} is fitted with the form
\begin{equation}
\label{eq:SMRS}
v_\pi(x) = A x^b (1-x)^c\ ,
\end{equation}
with the parameters $A$, $b$ and $c$ determined at an input scale of
$Q^2=4$~GeV$^2$, given in Table~\ref{tab:param}.
According to Regge theory, the parameter $b$, which controls the $x
\to 0$ behavior, is given by the intercept of the $\rho$ meson
trajectory, and is predicted to be $b \approx -1/2$.  The parameter
$c$ dictates the asymptotic behavior as $x \to 1$, and is predicted by
hadron helicity conservation in perturbative QCD to be $c=2$ for the
pion \cite{Farrar:yb,Lepage:1980fj,GUNION}.
The GRS (next-to-leading order) parameterization \cite{Gluck:1999xe}
contains two additional parameters,
\begin{equation}
\label{eq:GRS}
v_\pi(x) = A x^b (1-x)^c (1+e\sqrt{x}+g x)\ ,
\end{equation}
with all parameters listed for $Q^2=5$~GeV$^2$ in
Table~\ref{tab:param}.  The small differences in scale between the
parameterizations and the lattice moments are negligible.

The phenomenological valence moments with which the lattice
calculations are compared are defined by averaging the integrals of
these two distributions, and the errors are calculated as the
difference between the moments of the two distributions.  These
average moments are given in Table~\ref{tab:moment} and shown at the
physical pion mass as open stars (valence) and open triangles (total)
in Fig.~\ref{fig:xtrap}.

The Mellin transform of the (more general) parameterization in
Eq.~(\ref{eq:GRS}) is given by
\begin{eqnarray}
\label{eq:momfit}
\langle x^n \rangle_{\rm val}
&=& A \big[ \beta(1+c,1+b+n)\ +\ e\ \beta(1+c,3/2+b+n) \nonumber \\
& & \hspace*{1cm} + g\ \beta(1+c,2+b+n) \big]\ ,
\end{eqnarray}
where $\beta(a,b)$ is the $\beta$-function.  For the simpler SMRS
parameterization, only the first term in Eq.~(\ref{eq:momfit}) is
present.

To determine our ability to reconstruct the parameters of a
distribution from its moments, we first calculate the moments of the
GRS distribution (to be specific) by direct numerical integration.
Using Eq.~(\ref{eq:momfit}), we find that the five parameters in
Eq.~(\ref{eq:momfit}) can be very accurately reconstructed (to 4
significant figures) from the first five moments ($n=0-4$) using a
standard Levenberg-Marquardt non-linear fit.
However, since only 3 non-trivial lattice moments are currently
available, one cannot determine all of the five parameters from the
lattice data.
If we use the simpler form with $e=g=0$ in Eq.~(\ref{eq:momfit}) to
fit the lowest three non-trivial moments, the parameters $b$ and $c$
that give the best fit to the data differ from those of the underlying
distribution by approximately 10\% and 30\%, respectively.
This provides a guide to the size of systematic errors associated with
the choice of the parametric form. 

\subsection{Valence distribution from lattice moments}
\label{sec:valence}

Having investigated the accuracy with which one can reliably extract
the $x$ dependence of the valence quark distribution from the lowest
few moments, we now turn to the extrapolated lattice data discussed in
Section~\ref{sec:xtrap} and use these to fit the parameters $A$, $b$
and $c$ in Eq.~(\ref{eq:SMRS}).

There are several possible approaches to reconstructing the $x$
distribution from the available data, which we discuss in the
following.

(i) Ideally, the $n$-odd and $n$-even moments should be fitted
independently as they correspond to different distributions
(Eq.~(\ref{eq:moments})), and the valence distribution extracted from
the $n$-even ($C$-odd) lattice moments.  In this approach (which we
refer to as method (i)), both the statistical and systematic
uncertainties (associated with the fact that current lattice data do
not constrain $\mu$) of the various extrapolations would be improved
by future lattice data.  However, the two available ($n=0,2$) moments
are not sufficient to constrain all of the parameters in the standard
form of Eq.~(\ref{eq:SMRS}).  With the existing data, taking $\mu =
0.7$~GeV in the extrapolation, the $n=0$ moment fixes $A$, and we find
a minimum $\chi^2$ along the line $b \approx -0.9 + 0.2 c$ (for $0 < c
\alt 4$).
For the case $c=1$, one has $b \approx -0.7$, while for $c=2$, $b
\approx -0.5$.  Both of these curves are in qualitative agreement with
the Drell-Yan data.
If it proved feasible to extract the $n=4$ and 6 lattice moments in
the future, this method would be ideal.

(ii) An alternative approach is to assume, as discussed in
Sec.~\ref{sec:xtrap}, that quark loops are suppressed at large $m_\pi$
and that the $n$-odd valence moments are approximately equal to the
calculated $C$-even moments in that mass region.  This provides us
with 4 valence moments to which we fit 3 parameters.  We choose
$\mu=0.7$~GeV for the central extrapolation, taking $\mu=0.4$ and
1.0~GeV and the lattice data $\pm$ its quoted errors, respectively, as
a conservative measure of the overall error.  This yields the
parameters shown in Table~\ref{tab:param} (method (ii)).  We determine
the uncertainty in the parameters (arising from the statistical errors
in the lattice data and the systematic errors in the extrapolation
(choice of $\mu$)) by choosing an ensemble of sets of randomly
distributed moments within the extrapolated bounds and computing their
standard deviation.  As discussed above, there is additional
systematic uncertainty arising from the reconstruction procedure that
is not shown.

\begin{figure}[!t]
  \includegraphics[width=\columnwidth]{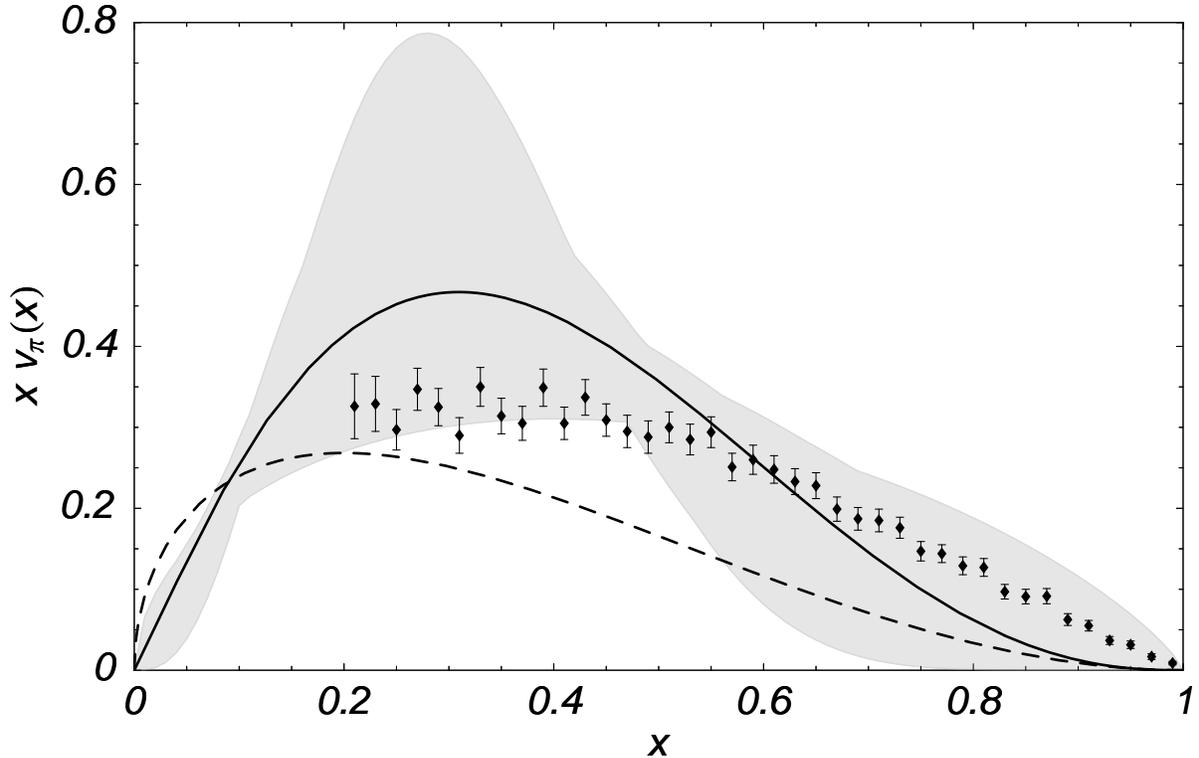}
\caption{\label{fig:x_nall} Valence distribution of the pion, 
  reconstructed using method (ii) described in the text. The shaded
  region corresponds to the uncertainty in the distribution
  (see text).  The dashed line corresponds to $v_\pi(x) =
  Ax^{-1/2}(1-x)^2$, which incorporates the hadron helicity
  conservation expectation for the large-$x$ behavior.  Experimental
  Drell-Yan data (diamonds) are from Ref.~\protect\cite{Conway:fs}.}
\end{figure}

The resulting $x$ distribution, which is illustrated in
Fig.~\ref{fig:x_nall}, is in qualitative agreement with the Drell-Yan
pion structure function data \cite{Conway:fs}.  At intermediate values
of $x$, the extracted distribution tends to lie a little above the
experimental data, while at large $x$ it appears to lie slightly below
-- with an $x \to 1$ behavior similar to that predicted by hadron
helicity conservation in perturbative QCD. Of course, given the
relatively large errors at present (the shaded region is the envelope
of the ensemble of reconstructed distributions used in the error
analysis), the distribution shows no significant disagreement with the
experimental data.

\begin{figure}[!t]
  \includegraphics[width=\columnwidth]{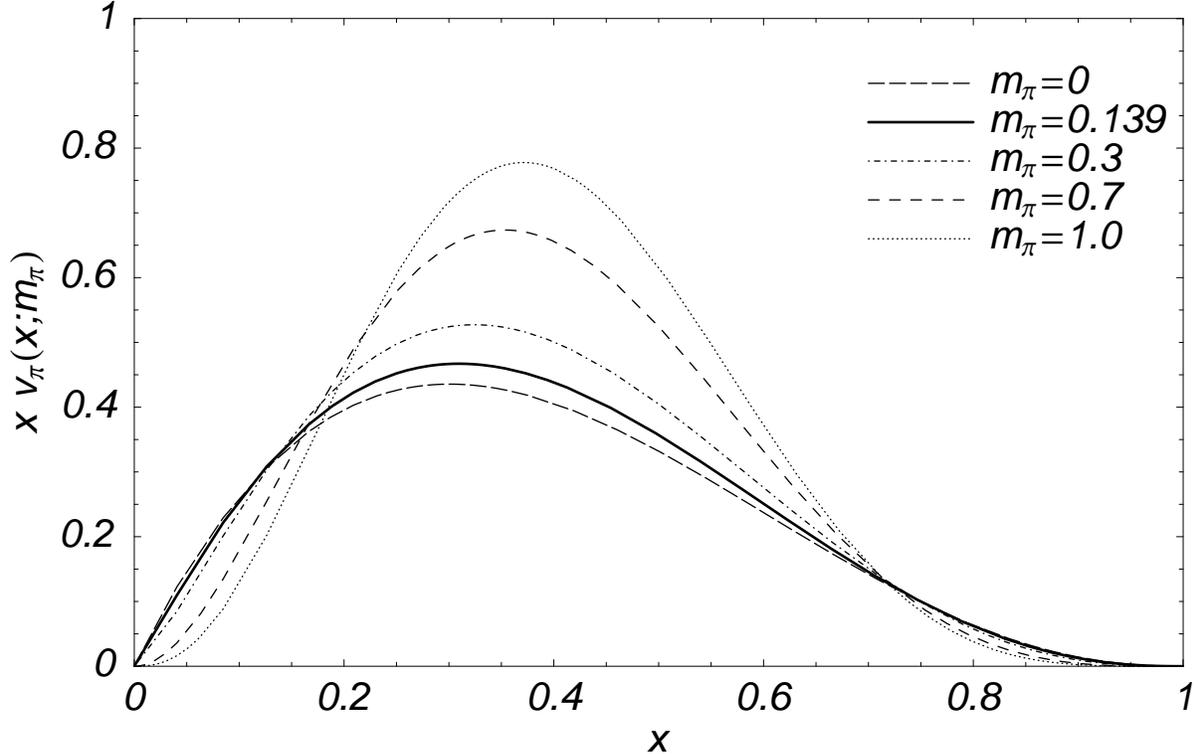}
\caption{\label{fig:massplot} Mass dependence 
  of the valence quark distribution of the pion (with $m_\pi$ in GeV).}
\end{figure}

Since Eq.~(\ref{eq:xtrap_NS}) gives the $m_\pi$ dependence of the
moments, we can examine the dependence of the valence distribution on
the pion mass. The result of reconstructing the PDF at several values
of $m_\pi^2$ is shown in Fig.~\ref{fig:massplot}. We do not show
results for values of $m_\pi^2$ larger than 1~GeV$^2$ because there
are no lattice data to constrain the reconstruction.  However, we have
checked that if the heavy quark limit is built into our extrapolation
function, the distribution approaches a $\delta$-function at $x=1/2$.
It is interesting that, even without such a constraint, the PDF seems
to show that the momentum of the pion is shared primarily between the
two valence quarks for $m_\pi$ above 0.7~GeV.

The obvious problem with this method is that the assumption that the
$C$-odd and $C$-even moments are approximately equal must break down
as the lattice data are extended to lower masses --- presumably where
one begins to see curvature in the $n=2$ moment.

(iii) A third possibility is to extrapolate the $n$-odd
moments linearly, according to Eq.~(\ref{eq:xtrap_S}), and subtract
twice the moments of the phenomenological sea at the physical mass to
give the valence moments.  The disadvantage of this method is that it
relies on phenomenological information in addition to that obtained
from the lattice.  Moreover, the sea quark distribution in the pion is
only very weakly constrained by experimental data, so that the errors
on the valence, $n$-odd moments will be large compared to those on the
$n$-odd total moments extracted from the lattice. A further
disadvantage of this method, from the purely theoretical point of
view, is that in this approach one obviously cannot study the
variation of the pion PDFs as a function of quark mass, as was done
for method (ii) above and for the nucleon in
Ref.~\cite{Detmold:2001dv}.  On the other hand, the errors on the
extrapolations can be improved systematically as the lattice data at
smaller quark masses become available and new experiments better
constrain the pion sea \cite{JLAB12,HERA}.
\begin{figure}[!t]
  \includegraphics[width=\columnwidth]{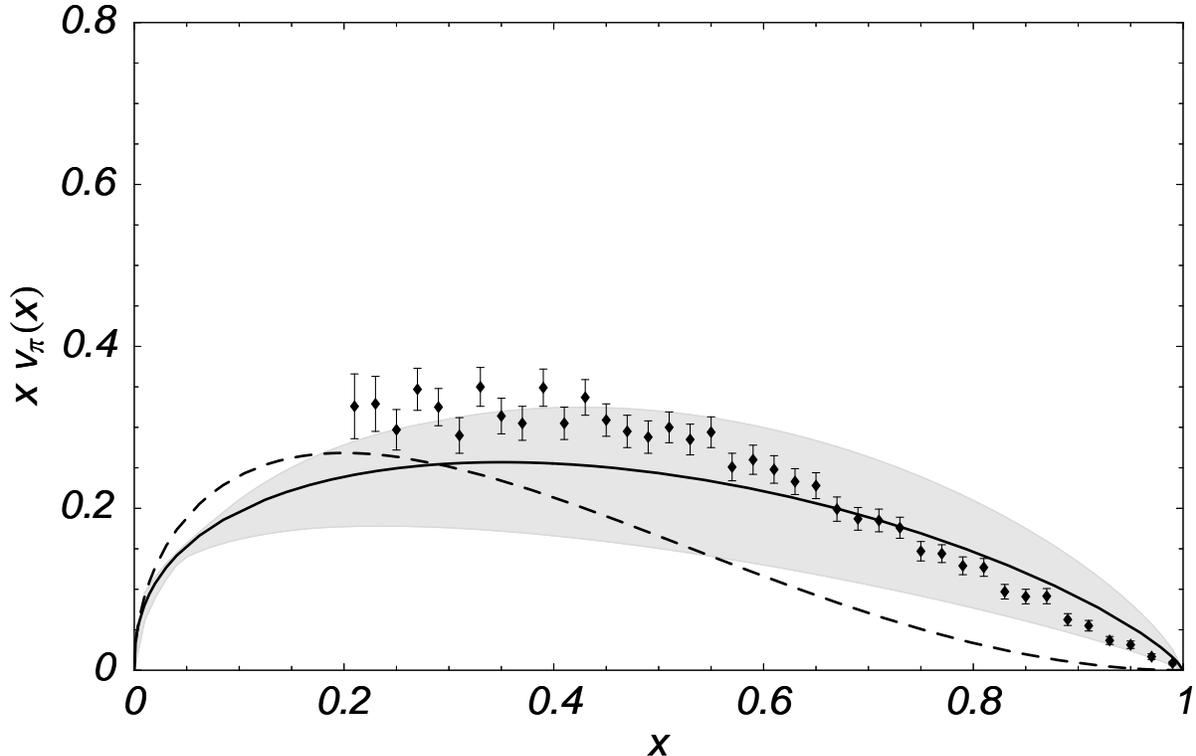}
\caption{\label{fig:x_neven} As in Fig.~\protect\ref{fig:x_nall},
  but using method (iii) to reconstruct the valence pion
  distribution.}
\end{figure}

Taking the $n$-odd sea moments from an average of the SMRS and GRS
distributions and extrapolating the $n=2$ moment with $\mu=0.7$~GeV
gives the parameters shown in Table~\ref{tab:param} as method (iii).
Errors are as described for method (ii) (where relevant).  These
parameters yield a distribution which is in good agreement with the
Drell-Yan data, as seen in Fig.~\ref{fig:x_neven}.  In particular, the
extracted curves are consistent with (though slightly harder than) the
$x \to 1$ behavior found in the experimental analyses, which, however,
disagree with the hadron helicity conservation predictions.

While there is some difference between the detailed $x$ dependence for
the valence quark distribution obtained using the methods (ii) and
(iii), these should disappear once more accurate lattice data on the
moments become available. Nevertheless, the fact that both methods
are in reasonable agreement with the Drell-Yan data, and with each
other, is very encouraging.  It would be particularly valuable to have
accurate higher moments ($n=4$--6) in order to constrain the detailed
shape of the distribution.
\begin{table}
\begin{ruledtabular}\begin{tabular}{cccccc}
Fit & $A$ & $b$ & $c$  \\
\hline
SMRS      & 1.08   & --0.36   & 1.08	\\
GRS       & 0.98   & --0.47   & 1.02	\\
method (ii)  & 4.4 & 0.1(5)   & 2.5(1.5)\\
method (iii) & 0.6 & --0.6(3) & 0.8(9)	\\
\end{tabular}\end{ruledtabular}
\vspace*{0.5cm}
\caption{\label{tab:param}
  Parameters of the valence distributions from various methods of
  analysis. The GRS valence
 distribution (\protect\ref{eq:GRS}) in addition uses the parameters
  $e=-0.81$ and $g=0.64$. The quoted errors combine the statistical
  and systematic (from $\mu$) errors. (Note that the lower limit on
  the parameter $c$ for method (iii) is constrained to be positive.)
  Errors are not given for $A$ as
  it is constrained by normalization. There is some additional systematic
  uncertainty from the reconstruction procedure which is not shown but
  is discussed in the text.}
\end{table}
%
 
\subsection{Sea distribution}
\label{sec:sea}

The sea quark distribution in the pion, defined as
\begin{eqnarray}
\label{eq:seadef}
s_\pi(x) &\equiv& {1\over 2}(q_\pi(x) - v_\pi(x))\ ,
\end{eqnarray}
is relatively poorly known experimentally.  There are no data from the
Drell-Yan reaction for $x \alt 0.2$ \cite{Betev:1985pg,Conway:fs}, and
the size of the sea is constrained only by imposing the momentum sum
rule.
A simple parameterization of the pion sea (as used by SMRS
\cite{Sutton:1991ay}) is
\begin{equation}
x\ s_\pi(x)= A_s (1-x)^\eta\ .
\label{eq:sea}
\end{equation}
The analysis of GRS \cite{Gluck:1999xe} determines the pion sea with
reference to the proton sea. A similar constraint can be derived at $x
\sim 10^{-2}$ from semi-inclusive DIS measurements at HERA, with the
result $F_2^\pi \approx F_2^N/3$ \cite{HERA}. This finding tends to
favor the SMRS sea over that of GRS \cite{Klasen:2001dd}, however,
this information corresponds to such small values of $x$ that it is of
little assistance for the present analysis.

In order to obtain information on the pion sea from the lattice data,
one would ideally extract the valence distribution according to method
(i) above, and use this to calculate the $n$-odd valence moments.
These would then be subtracted from the total moments, obtained from
the corresponding extrapolation of the lattice data to obtain the
$n$-odd moments of the pion sea at the physical pion mass. The $x$
dependence of the sea distribution could then be reconstructed using
the form (\ref{eq:sea}), given enough moments.

In the absence of sufficiently many moments for this to be a practical
solution, an alternative is to use the phenomenological valence ($n$-odd)
moments instead of the lattice moments.  Since these moments
are relatively well known, this procedure should be reasonably
reliable.  The $n$-odd sea moments which we extract using the linearly
extrapolated total moments from the lattice minus the phenomenological
valence moments are given in Table~\ref{tab:moment}. Using these data
to fit the Mellin transform of Eq.~(\ref{eq:sea}), we find the
parameters $A_s = 0.27$ and $\eta = 5.8$.
Unfortunately, the statistical uncertainty in these lattice sea
moments is large and the constraints on these parameters are very
weak.
In particular, the third moment of the sea is consistent with zero:
for $\langle x^3 \rangle_{\rm sea} \to 0$ the reconstruction gives
$\eta \to \infty$.
Nevertheless, this is in principle improvable and if the size of the
errors on the $n=3$ lattice moment were comparable to that on the
$n=1$ moment, a more robust reconstruction could be performed.

One could also modify this method by including the $n=2$ sea moment
constructed from the difference of the linear and chiral
extrapolations of the lattice data.  However, this introduces
additional uncertainty (from $\mu$) in the analysis and does not
reduce the uncertainty in the reconstructions.

\section{Summary and prospects}
\label{sec:summary}
 
We have studied the problem of the chiral extrapolation of the moments
of the PDFs of the pion, from the large quark masses where current
lattice QCD calculations are performed to the physical values. As in
earlier studies of the PDFs of the nucleon, the non-linearity of the
model independent non-analytic variation of the moments of the valence
PDF is extremely important, producing a significant change in the moments
at the physical quark mass, compared with a naive linear extrapolation.
In comparison, the moments of the singlet distribution, $q + \bar{q}$,
show no non-analytic behavior and are therefore expected to extrapolate
smoothly to the chiral limit.

Having studied the extrapolation of the moments of both the singlet
and non-singlet PDFs, we examine various procedures for reconstructing
the valence and sea distributions of the pion from the lattice
moments.  To make optimal use of the available lattice data for the
$n=1$--3 moments, we make the reasonable assumption that at large
quark masses ($m_\pi \agt 500$--600~MeV) the effects of quark loops
are suppressed, so that the effects of the quenched approximation and
disconnected insertions will not affect the extrapolation.
This allows the parameters of
the valence, and to some extent the sea, quark distributions to be
determined,
and the extracted distribution compared with the available Drell-Yan
data.  Over the range of intermediate $x$, from 0.2 to 0.8, the
reconstructed valence distribution is in fair agreement with the
Drell-Yan data, within the rather large errors arising from the
extrapolation procedure.  At this stage, however, it is not possible
to distinguish between the large-$x$ behavior predicted from hadron
helicity conservation in perturbative QCD and that found in the
Drell-Yan data.  Nevertheless, new lattice simulations, on the much
faster computers expected to be devoted to lattice QCD in the next few
years, should offer the chance, when analyzed using the techniques set
out here, to determine leading twist PDFs with an accuracy that
exceeds that of current experiments.

The results of this analysis can be used to guide future studies of
PDFs in lattice gauge theory.
Specifically,
\begin{itemize}
\item
The clearest observation is that it would be extremely valuable to
have quenched calculations for the $n=2$ and $n=3$ moments of an
accuracy comparable to that for $n=1$. This is especially important
for pion masses below 0.4~GeV$^2$. This alone would make a
substantial improvement in the errors on the parton distribution
functions. To better constrain the functional form of the $x$-dependence,
calculations of several higher moments (e.g. $n=4, 5$) would also be
desirable.
\item
In the case of the nucleon there is no observable difference between the
moments calculated in quenched and full QCD in the mass range covered.
It is vital to check that this is also true for the pion.
\item
In order to better constrain the extrapolation and to assure us that we
are on the right track it is important to push the lattice simulations to
lower pion masses. Ideally this should occur in full (unquenched) QCD;
however, quenched, and especially partially-quenched (which is a
computationally efficient way to get to smaller valence quark mass without
actually ignoring quark loops) simulations would also provide valuable
information to guide the chiral extrapolation.
\item
Of course, if one wants to explore the sea quark distribution, for which
there is little hard information at present, it will also be necessary to
include disconnected quark loops, even though it is difficult to pick out
a signal for such terms \cite{Lewis:2002tz}.
\item
Finally, as with all lattice simulations, we need to confirm that the
continuum ($a \rightarrow 0$) and infinite volume ($L \rightarrow \infty$)
corrections are fully under control.
\end{itemize}
With such a program we could expect significant advances over the next
few years in our understanding of the quark structure of the pion in QCD.

\acknowledgments{The authors are grateful to J.~Zanotti for helpful
  discussions. This work was supported by the Australian Research
  Council, the U.S. Department of Energy contract \mbox{DE-FG03-97ER41014}
  and contract \mbox{DE-AC05-84ER40150}, under which the Southeastern
  Universities Research Association (SURA) operates the Thomas
  Jefferson National Accelerator Facility (Jefferson Lab)}.

\end{document}